\documentclass[aps,pra,twocolumn,showpacs,superscriptaddress,floatfix]{revtex4}
\usepackage{graphicx,amsmath,amssymb}
\usepackage{times}
\usepackage{txfonts}

\begin{document}

\title{Collisional relaxation of Fesh\-bach molecules and three-body 
  recombination\\ in $^{87}$Rb Bose-Einstein condensates}

\author{G. Smirne}
\affiliation{Clarendon Laboratory, Department of Physics,
  University of Oxford, Parks Road, Oxford, OX1 3PU, UK}
\author{R.M. Godun}
\affiliation{Clarendon Laboratory, Department of Physics,
  University of Oxford, Parks Road, Oxford, OX1 3PU, UK}
\author{D. Cassettari}
\affiliation{Clarendon Laboratory, Department of Physics,
  University of Oxford, Parks Road, Oxford, OX1 3PU, UK}
%\affiliation{School of Physics and Astronomy, University of St.~Andrews, 
%  North Haugh, St.~Andrews, Fife, KY16 9SS, UK}
\author{V. Boyer}
\affiliation{Clarendon Laboratory, Department of Physics,
  University of Oxford, Parks Road, Oxford, OX1 3PU, UK}
\author{C.J. Foot}
\affiliation{Clarendon Laboratory, Department of Physics,
  University of Oxford, Parks Road, Oxford, OX1 3PU, UK}
\author{T. Volz}
\affiliation{Max-Planck-Institut f\"ur Quantenoptik, 
  Hans-Kopfermann-Strasse 1, 
  85748 Garching, Germany}
\author{N. Syassen}
\affiliation{Max-Planck-Institut f\"ur Quantenoptik, 
  Hans-Kopfermann-Strasse 1, 
  85748 Garching, Germany}
\author{S. D\"urr}
\affiliation{Max-Planck-Institut f\"ur Quantenoptik, 
  Hans-Kopfermann-Strasse 1, 
  85748 Garching, Germany}
\author{G. Rempe}
\affiliation{Max-Planck-Institut f\"ur Quantenoptik, 
  Hans-Kopfermann-Strasse 1, 
  85748 Garching, Germany}
\author{M.D. Lee}
\affiliation{Clarendon Laboratory, Department of Physics,
  University of Oxford, Parks Road, Oxford, OX1 3PU, UK}
\author{K. G{\'o}ral}
\affiliation{Clarendon Laboratory, Department of Physics,
  University of Oxford, Parks Road, Oxford, OX1 3PU, UK}
\author{T. K{\"o}hler}
\affiliation{Clarendon Laboratory, Department of Physics,
  University of Oxford, Parks Road, Oxford, OX1 3PU, UK}

\begin{abstract}
  We predict the resonance enhanced magnetic field dependence of atom-dimer 
  relaxation and three-body recombination rates in a $^{87}$Rb Bose-Einstein 
  condensate (BEC) close to 1007\,G. Our exact treatments of three-particle 
  scattering explicitly include the dependence of the interactions on the 
  atomic Zeeman levels. The Fesh\-bach resonance distorts the entire diatomic 
  energy spectrum causing interferences in both loss phenomena. Our two 
  independent experiments confirm the predicted recombination loss over a 
  range of rate constants that spans four orders of magnitude.
\end{abstract}

\date{\today}
\pacs{34.50.-s,03.75.-b,34.10.+x,21.45.+v}
\maketitle

Few-body collisions determine the lifetimes of cold gases of atoms and 
diatomic molecules. Recent experiments on identical Bose atoms 
\cite{ChinPRL05,KraemerNature06} suggest that magnetically tunable 
interactions \cite{InouyeNature98,RobertsPRL00} in combination with such 
scattering phenomena could allow association of atomic clusters at rest 
\cite{StollPRA05} and confirmation of long standing predictions of quantum 
physics \cite{EfimovPhysLett70}. Destructive interferences of inelastic 
three-body collisions have paved the way for the Bose-Einstein condensation of 
$^{133}$Cs \cite{KraemerNature06}. Scattering processes involving three atoms 
are known as three-body recombination and atom-dimer collisions. Three-body 
recombination refers to the threshold-less transition from initially unbound 
atoms to a dimer molecule and a remnant atom in accordance with energy and 
momentum conservation. All three atoms are lost from a cold gas. Previous 
theoretical studies relied upon model calculations \cite{Efimovresonances} or 
analytic estimates \cite{PetrovPRL04}, suggesting general trends for the 
behaviour of recombination rates in the limit of large diatomic scattering 
lengths. Several {\em ab initio} approaches to cold atom-dimer collisions 
\cite{abinitiorelaxation} are presently limited to comparatively tightly bound 
diatomic states with energies up to about $-k_\mathrm{B}\times 1\,$K 
($k_\mathrm{B}=1.38\times 10^{-23}$\,J/K). The energies of dimers produced via 
magnetically tunable interactions are, however, typically on the order of 
$-k_\mathrm{B}\times 1\,$mK, right below the dissociation threshold. 

In this paper we predict, for the first time, the atom-dimer collision rates 
involving such Fesh\-bach molecules, consisting here of $^{87}$Rb atoms, 
without any adjustable parameters. Our exact approach to the three-body 
Schr\"odinger equation explicitly accounts for the coupling between the atomic 
Zeeman states. This allows us to properly describe the magnetic field 
dependent distortion of the diatomic bound state energies 
\cite{DuerrPRL04,Koehlercondmat06} caused by the weakly coupled Fesh\-bach 
resonance \cite{MartePRL02,VolzPRA03}. We predict constructive and destructive 
interferences in both atom-dimer relaxation and three-body recombination rates 
which we interpret in terms of resonance phenomena associated with meta-stable 
trimer or bound dimer levels. Our approach has been tested by two independent 
experiments on three-body recombination in a $^{87}$Rb BEC over a range of 
rate constants that spans four orders of magnitude.

The three-body Hamiltonian, $H_\mathrm{3B}$, is comprised of the kinetic 
energies of the atoms and their pairwise interactions, 
i.e.~$H_\mathrm{3B}=H_0+V_1+V_2+V_3$. Here $V_1$ is the potential associated 
with the interaction of the atom pair $(2,3)$, whereas $V_2$ and $V_3$ follow 
from cyclic permutations of the indices. We neglect short range genuinely 
three-body interactions as they are not expected to significantly affect 
resonance enhanced collisions. In the barycentric frame the total kinetic 
energy, $H_0$, can be divided into contributions from the relative motion of 
an atom pair and the motion of the third atom with respect to the centre of 
mass of the pair. Given the associated spatial Jacobi coordinates, 
$\mathbf{r}$ and $\boldsymbol{\rho}$, respectively, such a separation yields: 
\begin{equation}
  H_0=|\mathrm{bg}\rangle
  \left(
  -\frac{3\hbar^2}{4m}
  {\boldsymbol{\nabla}}_{\boldsymbol{\rho}}^2
  -\frac{\hbar^2}{m}
  {\boldsymbol{\nabla}}_{\mathbf{r}}^2
  \right)
  \langle\mathrm{bg}|.
\end{equation}
Here $m$ is the atomic mass and $|\mathrm{bg}\rangle$ indicates a product of 
$(f=1,m_f=1)$ Zeeman states \cite{Koehlercondmat06} in which the three atoms 
of the $^{87}$Rb BEC are prepared. Depending on the context, we refer to 
either a product of three or two of such atomic ground states as the 
entrance-channel spin configuration and choose the zero of energy at its 
threshold for dissociation into unbound atoms.

To describe their initial and final configurations, we label each one of the 
colliding atoms by a Greek index $\alpha=1,2,3$. The pairwise interactions, 
$V_\alpha$, determine the properties of cold diatomic collisions and highly 
excited molecular bound states, as well as their dependence on the magnetic 
field strength, $B$. In an idealised treatment, the observed resonant 
enhancement of the binary scattering length in the vicinity of $B_0=1007.4$\,G 
\cite{MartePRL02,VolzPRA03} arises from the near degeneracy of a single 
meta-stable diatomic energy level with the entrance-channel dissociation 
threshold. This bare Fesh\-bach resonance level, $|\phi_\mathrm{res}\rangle$, 
refers to a closed scattering channel comprised of Zeeman states from the 
excited $f=2$ level \cite{Koehlercondmat06}. Accordingly, the interactions can 
be represented by \cite{Koehlercondmat06} 
\begin{align}
  \nonumber
  V_\alpha=&|\mathrm{bg}\rangle V_\alpha^\mathrm{bg} \langle\mathrm{bg}|
  +W_\alpha |\phi_\mathrm{res}\rangle_\alpha\, 
  {_\alpha}\langle\phi_\mathrm{res}|\\
  &+|\phi_\mathrm{res}\rangle_\alpha\, 
  {_\alpha}\langle\phi_\mathrm{res}|W_\alpha
  +|\phi_\mathrm{res}\rangle_\alpha\,E_\mathrm{res}(B)\, 
  {_\alpha}\langle\phi_\mathrm{res}|.
  \label{potentials}	
\end{align}
Here $V_\alpha^\mathrm{bg}$ and $|\phi_\mathrm{res}\rangle_\alpha$ are the 
diatomic entrance-channel potential and resonance level, respectively, whereas 
the third atom with index $\alpha$ is in the Zeeman ground state and plays the 
role of a spectator. Similarly, $W_\alpha$ describes the pairwise 
inter-channel spin exchange coupling. The resonance state energy, 
$E_\mathrm{res}$, depends on $B$ through
$E_\mathrm{res}(B)=\mu_\mathrm{res}(B-B_\mathrm{res})$. Here 
$\mu_\mathrm{res}=h\times 4.2\,$MHz/G is the difference in magnetic moments of 
closed- and entrance-channel atom pairs \cite{DuerrPRL04}, and 
$B_\mathrm{res}$ indicates the point of degeneracy of $E_\mathrm{res}$ with 
the entrance-channel dissociation threshold. As the resonance state is weakly 
coupled, the shift $B_0-B_\mathrm{res}$ is negligible \cite{Koehlercondmat06}. 
Due to the absence of overlap between Zeeman states constituting the diatomic 
entrance ($f=1$) and closed ($f=2$) channels, all resonance states associated 
with different arrangements of the atoms are orthogonal:
${_\alpha}\langle\phi_\mathrm{res}|\phi_\mathrm{res}\rangle_\beta=
\delta_{\alpha\beta}$.

Our exact solutions of the three-body Schr\"odinger equation are based on the 
Alt, Grassberger, and Sandhas (AGS) technique \cite{AltNuclPhysB67}. 
Accordingly, we introduce the complete Green's function, 
$G_\mathrm{3B}(z)=(z-H_\mathrm{3B})^{-1}$, which characterises all bound and 
continuum levels of $H_\mathrm{3B}$. Here $z=E_\mathrm{i}+i0$ is a complex 
variable indicating that the physical continuum energy of incoming particles, 
$E_\mathrm{i}$, is approached from the upper half of the complex plane. Given 
the arrangement-channel and free Green's functions, 
$G_\alpha(z)=(z-H_0-V_\alpha)^{-1}$ and $G_0(z)=(z-H_0)^{-1}$, respectively, 
the AGS transition matrices, $U_{\alpha\beta}(z)$, are defined by implicit 
relations ($\alpha,\beta=1,2,3$):
\begin{equation}
  G_\mathrm{3B}(z)=\delta_{\alpha\beta}G_\beta(z)
  +G_\alpha(z)U_{\alpha\beta}(z)G_\beta(z).
  \label{AGSdecomposition}
\end{equation}
In accordance with Ref.~\cite{AltNuclPhysB67}, these transition matrices 
determine, e.g., the Bose-symmetric scattering amplitudes for atom-dimer 
elastic or relaxation collisions by
\begin{equation}
  f(\mathbf{q}_\mathrm{f},\mathbf{q}_\mathrm{i})
  =-\frac{8\pi^2m\hbar}{9}\sum_{\alpha,\beta=1}^3
  {_\alpha}\langle\mathbf{q}_\mathrm{f},\phi_\mathrm{b}^\mathrm{f}|
  U_{\alpha\beta}(z)
  |\mathbf{q}_\mathrm{i},\phi_\mathrm{b}^\mathrm{i}\rangle_\beta. 
  \label{amplitudes} 
\end{equation}
Here $|\mathbf{q}_\mathrm{i},\phi_\mathrm{b}^\mathrm{i}\rangle_\beta$ 
indicates an incoming dimer in the coupled-channels state 
$|\phi_\mathrm{b}^\mathrm{i}\rangle$ of energy $E_\mathrm{b}^\mathrm{i}$, 
whereas the third atom with index $\beta$ has the momentum 
$\mathbf{q}_\mathrm{i}$ with respect to the molecular centre of mass, and is 
described by 
$\langle\boldsymbol{\rho}|\mathbf{q}_\mathrm{i}\rangle=
\exp(i\mathbf{q}_\mathrm{i}\cdot\boldsymbol{\rho}/\hbar)/(2\pi\hbar)^{3/2}$.
Consequently, the initial state fulfils 
$(H_0+V_\beta)|\mathbf{q}_\mathrm{i},\phi_\mathrm{b}^\mathrm{i}\rangle_\beta
=E_\mathrm{i}|\mathbf{q}_\mathrm{i},\phi_\mathrm{b}^\mathrm{i}\rangle_\beta$
with $E_\mathrm{i}=3q_\mathrm{i}^2/(4m)+E_\mathrm{b}^\mathrm{i}$.
Similarly, $|\mathbf{q}_\mathrm{f},\phi_\mathrm{b}^\mathrm{f}\rangle_\alpha$ 
describes the scattered atom and dimer, whose final energy,
$E_\mathrm{f}=3q_\mathrm{f}^2/(4m)+E_\mathrm{b}^\mathrm{f}$, equals 
$E_\mathrm{i}$. The transition matrices in Eq.~(\ref{amplitudes}) are 
determined by the AGS equations \cite{AltNuclPhysB67}:
\begin{equation}
  U_{\alpha\beta}(z)=(1-\delta_{\alpha\beta})G_0^{-1}(z)
  +\sum_{{\gamma=1}\atop{\gamma\neq\alpha}}^{3}
  V_\gamma G_\gamma(z)U_{\gamma\beta}(z).
  \label{AGS}
\end{equation}

\begin{figure}[tbp]
  \includegraphics[width=\columnwidth,clip]{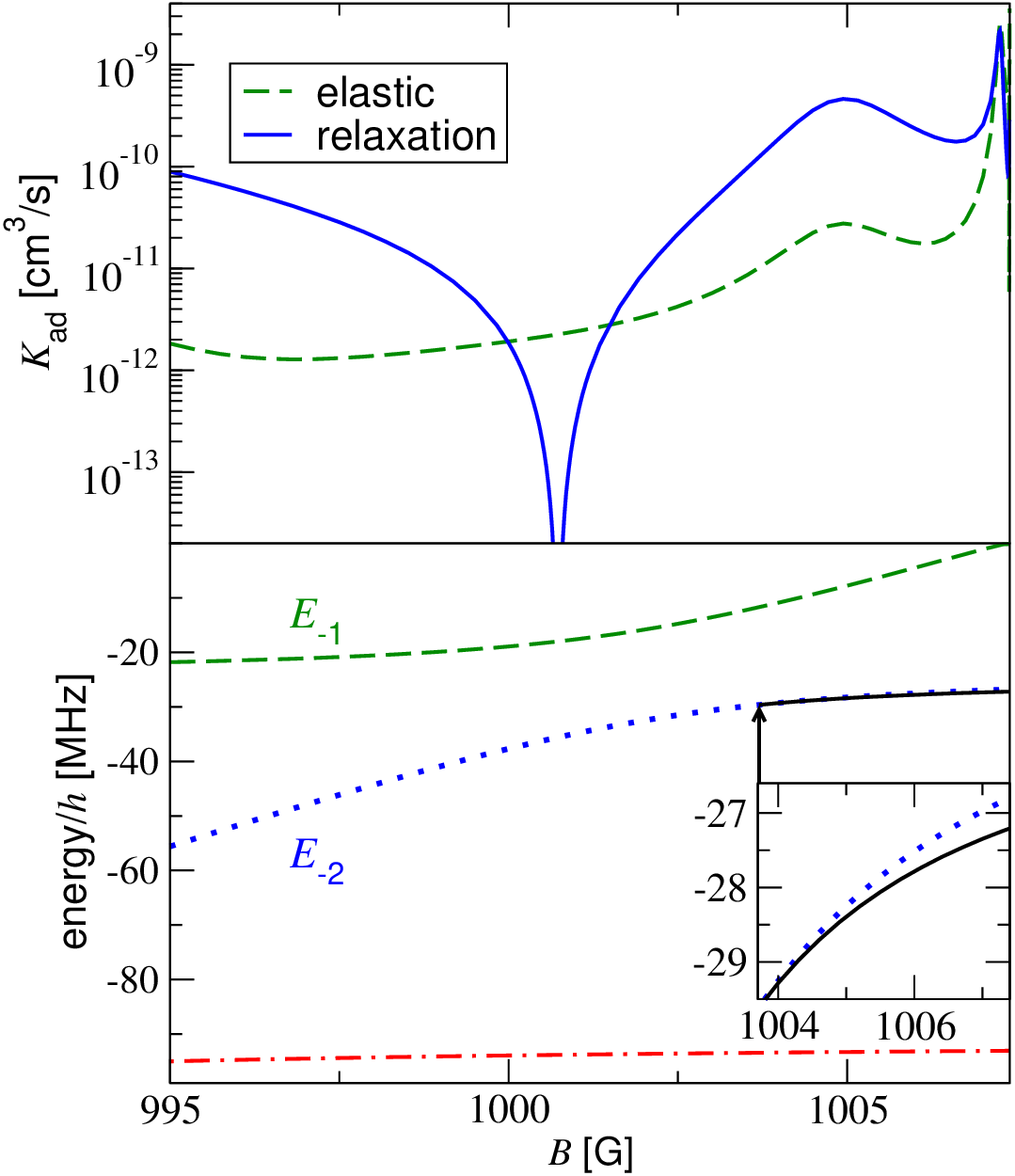}
  \caption{(Colour online)
    Upper panel: Elastic (dashed curve) and inelastic (solid curve) 
    atom-dimer scattering rates, $K_\mathrm{ad}$, associated with a collision 
    energy of $3q_\mathrm{i}^2/(4m)=k_\mathrm{B}\times 10$\,nK
    ($k_\mathrm{B}\times 1\,$mK$=h\times 20.8\,$MHz) versus $B$ on the 
    low-field side of $B_0$. Lower panel: Energies of the Fesh\-bach 
    molecular (dashed curve) and second highest excited vibrational (dotted 
    curve) dimer states, $E_{-1}$ and $E_{-2}$, respectively. The dot-dashed 
    and solid curves indicate the energies of isotropic meta-stable 
    $^{87}$Rb$_3$ states. The highest excited trimer level emerges at about 
    1003.7\,G (arrow) in a zero energy resonance (inset) for collisions 
    between atoms and dimers in the second highest excited vibrational state.}
  \label{fig:relaxation}
\end{figure}

Based on Eq.~(\ref{AGS}), the upper panel of Fig.~\ref{fig:relaxation} 
illustrates the rate constants, 
$K_\mathrm{ad}=\sigma(q_\mathrm{f}) q_\mathrm{f}/(2m/3)$, for elastic 
scattering ($E_\mathrm{b}^\mathrm{f}=E_\mathrm{b}^\mathrm{i}=E_{-1}$) as 
well as relaxation ($E_\mathrm{b}^\mathrm{f}<E_\mathrm{b}^\mathrm{i}=E_{-1}$) 
of $^{87}$Rb$_2$ Fesh\-bach molecules. Here $\sigma(q_\mathrm{f})$ denotes 
the (in)elastic $s$-wave cross section associated with the amplitude of 
Eq.~(\ref{amplitudes}). Our implementation of Eq.~(\ref{potentials}) follows 
the approach of Ref.~\cite{GoralJPhysB04}. Consequently, the interactions 
properly describe: the measured $B$-dependences of the diatomic scattering 
length \cite{VolzPRA03,DuerrPRA04}, the binding energies of the two highest 
excited vibrational $^{87}$Rb$_2$ states \cite{DuerrPRL04}, their long-range 
wave functions as well as their entrance- and closed-channel spin admixtures 
\cite{Koehlercondmat06}. The next vibrational dimer level is far detuned from 
the dissociation threshold by about 600\,MHz \cite{GoralJPhysB04} and will be 
neglected. Our predicted collisional relaxation rate constants are typically 
on the order of $10^{-10}\,$cm$^3$/s. Such magnitudes have been confirmed by 
experiments on the stability of $^{87}$Rb$_2$, yielding $K_\mathrm{ad}=2\times 
10^{-10}\,$cm$^3/$s at 1005.8\,G \cite{Syassencondmat06} in good agreement 
with Fig.~\ref{fig:relaxation}, as well as for $^{23}$Na$_2$ Fesh\-bach 
molecules \cite{MukaiyamaPRL04}. Both species are associated with weakly 
coupled, closed-channel dominated \cite{Koehlercondmat06}, resonances. The 
fast atom-dimer decay tends to exceed elastic scattering, except for a region 
centred at about 1001\,G, close to the avoided crossing of $E_{-1}$ and 
$E_{-2}$ at $1001.7\,$G \cite{DuerrPRL04} in the lower panel of 
Fig.~\ref{fig:relaxation}. We expect relaxation into more tightly bound dimer 
states to partly fill in the gap in the inelastic rate constants.

Three-body recombination in a BEC is well described in terms of the transition 
from a continuum momentum state, $|0,\mathrm{bg}\rangle$, associated with 
separated atoms at rest, to a dimer and a remnant atom. Whereas the initial 
state thus fulfils $H_0|0,\mathrm{bg}\rangle=0$, the final states are of the 
same nature as those in atom-dimer relaxation and elastic collisions. In 
accordance with Ref.~\cite{AltNuclPhysB67}, the associated transition matrix 
is given by $U_{\alpha,\beta=0}(z)$ of Eq.~(\ref{AGSdecomposition}) with 
$z=i0$. For a definite target dimer state $|\phi_\mathrm{b}^\mathrm{f}\rangle$ 
of energy $E_\mathrm{b}^\mathrm{f}$ this matrix gives the loss rate constant 
to be:   
\begin{equation}
  K_3=\frac{4\pi mq_\mathrm{f}(2\pi\hbar)^6}{\hbar}
  \sum_{\alpha=1}^3\int d\Omega
  \left|
      {_\alpha}\langle\mathbf{q}_\mathrm{f},\phi_\mathrm{b}^\mathrm{f}|
      U_{\alpha 0}(i0)
      |0,\mathrm{bg}\rangle
      \right|^2.
      \label{K3}
\end{equation}
Here $d\Omega$ refers to the angular component of $\mathbf{q}_\mathrm{f}$ 
whose modulus, $q_\mathrm{f}$, is determined by  
$3q_\mathrm{f}^2/(4m)+E_\mathrm{b}^\mathrm{f}=0$. The total loss rate 
constant, $K_3^\mathrm{tot}$, is found by summation of Eq.~(\ref{K3}) over all 
target states. Consequently, the number of condensate atoms, $N$, decays in 
accordance with  
\begin{equation}
  \dot{N}=-K_3^\mathrm{tot}\langle n^2\rangle N/6. 
  \label{recombinationdynamics}
\end{equation}
Here $\langle n^2\rangle$ is the mean square density and the factor $1/6$ 
accounts for the coherent nature of the gas \cite{factor6}. The transition 
matrix, $U_{\alpha 0}(i0)$, can be inferred from the solutions of the AGS 
equations \cite{AltNuclPhysB67} using $\beta=0$ in Eq.~(\ref{AGS}).

\begin{figure}[htbp]
  \includegraphics[width=\columnwidth,clip]{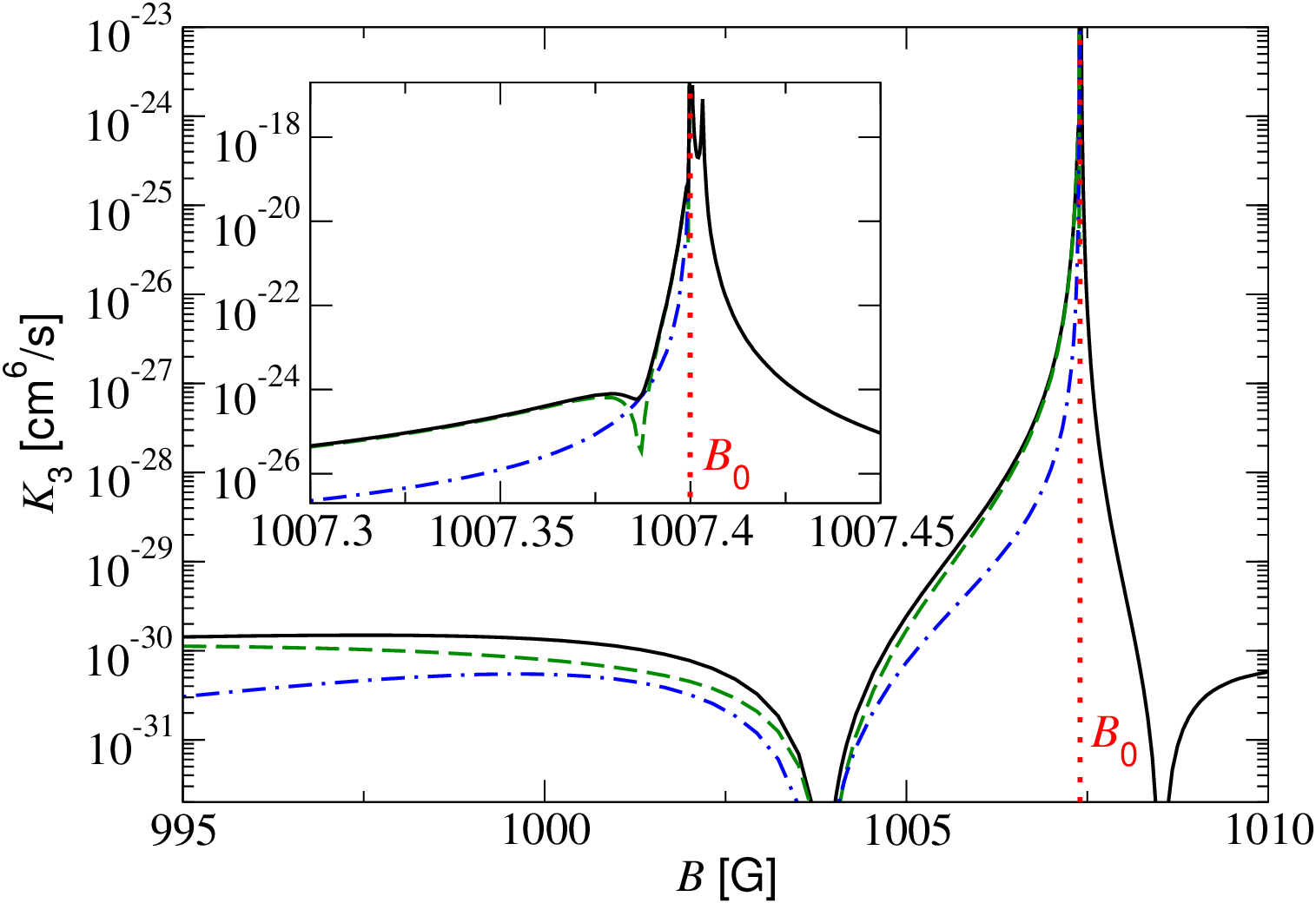}
  \caption{(Colour online) 
    Three-body recombination loss rate constant, $K_3$, versus $B$. Solid 
    curves refer to the total loss rate constant, $K_3^\mathrm{tot}$, whereas 
    dashed and dot-dashed curves are its contributions from the final 
    Fesh\-bach molecular and second highest excited vibrational $^{87}$Rb$_2$ 
    states, respectively. The inset shows an enlargement close to the diatomic 
    resonance whose position, $B_0$, is indicated by the dotted line.}
  \label{fig:K3overview}
\end{figure}

Figure~\ref{fig:K3overview} gives an overview of the predicted total loss rate 
constants, $K_3^\mathrm{tot}$, as well as their contributions from the 
different dimer target states. Recombination into Fesh\-bach molecules 
dominates $K_3^\mathrm{tot}$ typically by an order of magnitude. Similarly to
the atom-dimer collision rate constants of Fig.~\ref{fig:relaxation}, the 
loss rate constant $K_3^\mathrm{tot}$ shows interference minima which may be 
attributed to $B$-dependent changes in the bound states of $H_\mathrm{3B}$. 
Such isotropic trimer levels are indicated by the solid and dot-dashed curves 
in the lower panel of Fig.~\ref{fig:relaxation} and were determined using the 
Faddeev approach of Ref.~\cite{StollPRA05}. The minimum of $K_3^\mathrm{tot}$ 
at about 1004\,G in Fig.~\ref{fig:K3overview} occurs close to an atom-dimer 
zero energy resonance indicated by the arrow in Fig.~\ref{fig:relaxation}. At 
the resonant magnetic field strength the trimer level becomes degenerate with 
the second highest excited diatomic vibrational state where it can decay into 
a dimer and an atom in accordance with energy conservation. Both trimer levels 
of Fig.~\ref{fig:relaxation} are unstable with respect to such decay processes 
involving more tightly bound dimer states with energies below 
$-h\times 600\,$MHz. In the inset of Fig.~\ref{fig:K3overview} the 
interference minimum below $B_0$ and the maximum just above $B_0$ can be 
attributed respectively to atom-dimer and three-body zero energy resonances 
associated with meta-stable Thomas-Efimov trimer levels 
\cite{KraemerNature06,Efimovresonances,StollPRA05}. Similarly, the 
oscillations in the atom-dimer collision rates close to $B_0$ in 
Fig.~\ref{fig:relaxation} result from Efimov's effect \cite{EfimovPhysLett70}, 
but their narrow magnetic field range makes this physics largely inaccessible 
to experiments on $^{87}$Rb. The minima of $K_3^\mathrm{tot}$ at about 1004\,G 
and 1008.5\,G in Fig.~\ref{fig:K3overview}, however, could be resolved. 

\begin{figure}[htbp]
  \includegraphics[width=\columnwidth,clip]{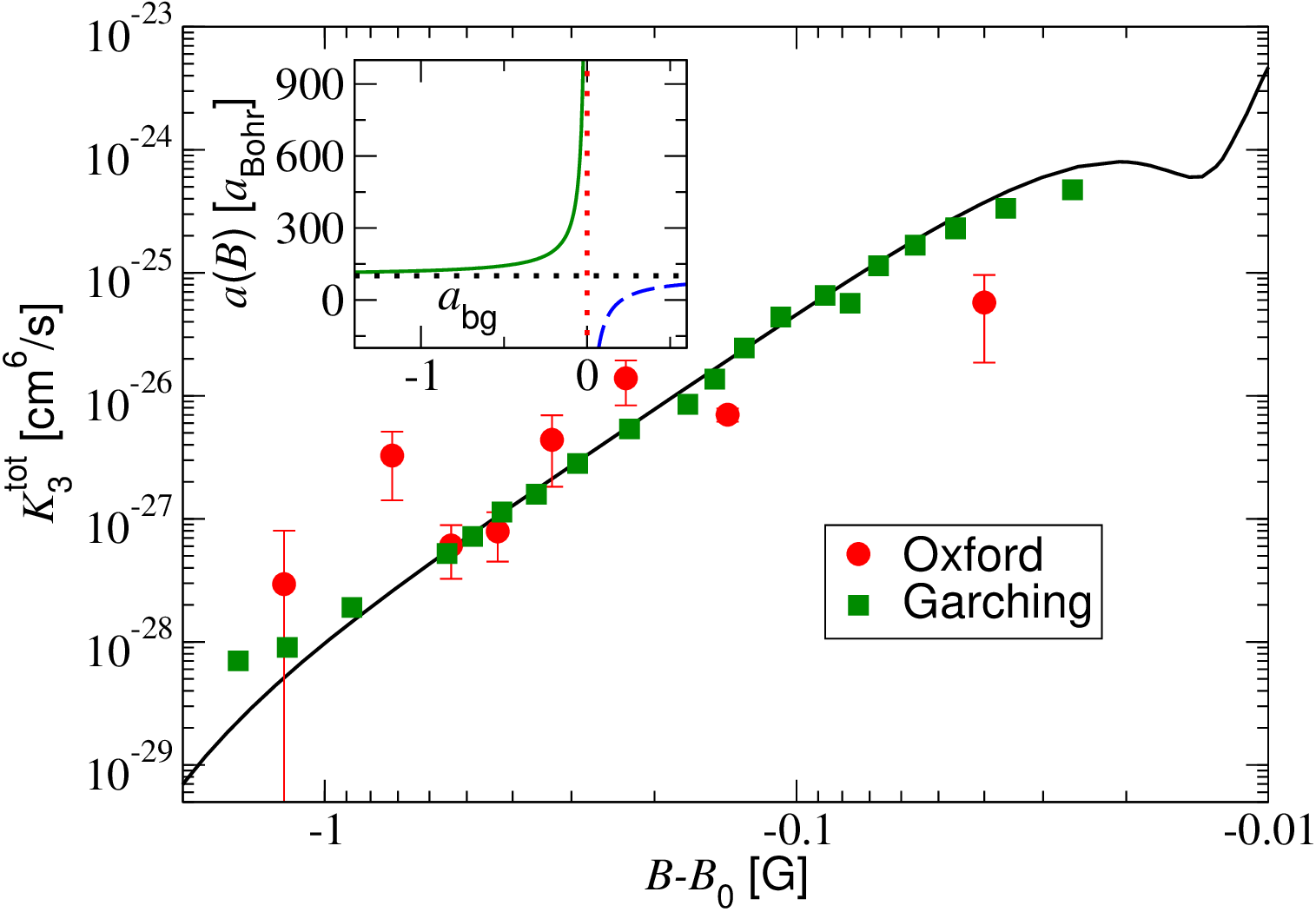}
  \caption{(Colour online) 
    Comparison between measured (circles and squares) and predicted 
    (solid curve) total three-body recombination loss rate constants, 
    $K_3^\mathrm{tot}$. Error bars refer to statistical uncertainties only.
    The inset shows the binary scattering length, 
    $a(B)=a_\mathrm{bg}[1-\Delta B/(B-B_0)]$, on the low-field (solid curve) 
    and high-field (dashed curve) sides of the diatomic zero energy resonance 
    (vertical dotted line). The horizontal dotted line indicates the 
    background scattering length \cite{VolzPRA03,DuerrPRA04} 
    $a_\mathrm{bg}=100.5\,a_\mathrm{Bohr}$ ($a_\mathrm{Bohr}=0.0529\,$nm)
    whereas the resonance width is $\Delta B=0.21\,$G. All measurements refer 
    to $B<B_0$ where the BEC is stable with respect to collapse.}
  \label{fig:K3exp}
\end{figure}

Figure~\ref{fig:K3exp} compares the predicted $K_3^\mathrm{tot}$ as a function 
of $B$ with two independent experiments. In accordance with the inset, the 
measurements cover a range of diatomic scattering lengths from about 100 to 
1000\,$a_\mathrm{Bohr}$ within which $K_3^\mathrm{tot}$ varies over four 
orders of magnitude. In the Oxford experiment, indicated by circles in 
Fig.~\ref{fig:K3exp}, $^{87}$Rb atoms are loaded from a BEC into about 60 
wells of a one dimensional optical lattice. The lattice is formed from a 
retro-reflected, vertical, 850\,nm laser beam giving rise to individual wells 
with trapping frequencies $(47,47,1.75 \times 10^4)$\,Hz. The atoms are then 
prepared in the $(f=1,m_f=1)$ Zeeman ground state and a homogeneous magnetic 
field is ramped to a value $B$ just below $B_0$. At this stage the atom number 
is $N_0=3 \times 10^5$ and the mean square density is 
$\langle n_0^2\rangle =2 \times 10^{28}$\,cm$^{-6}$.  This is obtained by 
calculating the mean square density in each well using the BEC wave function 
from Ref.~\cite{PedriPRL01} and averaging over the populated sites. After a 
hold time of $\Delta t = 100$\,ms, the magnetic field is ramped off and the 
final atom number $N_\mathrm{f}$ is measured. Assuming that coherence is 
maintained and that the volume of the cloud used to calculate the density $n$ 
in Eq.~(\ref{recombinationdynamics}) remains unchanged, the loss rate constant 
is extracted from $K_3^\mathrm{tot}=6\left[ (N_0/N_\mathrm{f})^2-1 \right]/
(2\langle n_0^2\rangle\Delta t)$. Uncertainties in $N_0$ prevent measurements 
of $K_3^\mathrm{tot}$ below $\sim 5 \times 10^{-28}$\,cm$^6$/s. We have 
confirmed this analysis by dynamical simulations of the experimental sequence
using the non-linear Schr{\"o}dinger equation with a three-body loss term.

In the Garching experiment, indicated by squares in Fig.~\ref{fig:K3exp}, a 
BEC of $5\times 10^5$ atoms is loaded into a crossed-beam optical dipole trap 
%with a geometric mean of frequencies of 108\,Hz 
\cite{DuerrPRA04}. After preparing the gas far away from $B_0$ the magnetic 
field is ramped and held at its final value, $B$. For $B<1007.2\,$G the size 
of the cloud changes adiabatically during the experimental sequence. Using the 
Thomas-Fermi prediction for the mean square density, $K_3^\mathrm{tot}$ is 
obtained from a fit of Eq.~(\ref{recombinationdynamics}) to the loss data 
\cite{SoedingApplPhysB99}. For $B>1007.2\,$G the depletion is too fast for the 
BEC to adiabatically adapt its size. Consequently, we jump the magnetic field 
from the initial to its final value and measure the loss only on time scales 
sufficiently short for the cloud to keep its initial size. Systematic errors 
arise mainly from the calibration of the trap frequencies and the atom number, 
leading to an estimated uncertainty of a factor of 3 in $K_3^\mathrm{tot}$. 
Statistical errors are negligible in comparison. The accuracy of magnetic 
field calibration is 30\,mG. We believe that our $B$-fields are sufficiently 
far from $B_0$ for atom loss caused by collisional avalanches 
\cite{SchusterPRL01} to be negligible. As the BEC dynamics in the crossed-beam 
dipole trap is comparatively simple, the associated overall systematic 
uncertainty is smaller than in the optical lattice setup. The agreement within 
uncertainties between predicted and measured $K_3^\mathrm{tot}$ of both 
experiments in Fig.~\ref{fig:K3exp} confirms our approach. 

We have shown how spin-dependent potentials can be included in a practical 
and exact treatment of three-body scattering phenomena. Destructive 
interferences predicted in atom-dimer relaxation rates allow elastic 
scattering of atoms from generally rather unstable $^{87}$Rb$_2$ Fesh\-bach 
molecules. The predicted magnetic field dependent suppressions of three-body 
recombination can be used to increase the lifetimes of cold gases 
\cite{KraemerNature06}. While we have confirmed our approach by experiments on 
$^{87}$Rb condensates, it is applicable to all species, bosons and fermions, 
subject to closed- or entrance-channel dominated resonances 
\cite{Koehlercondmat06}.

This research has been supported by the Royal Society, the UK EPSRC, the EC 
Marie-Curie program, and the Cold Quantum Gases Network.

\end{document}